# Gravitational scaling in Beijing Subway Network

Biao Leng[1], Yali Cui[2], Jianyuan Wang[1], Zhang Xiong[1], Shlomo Havlin[3], Daqing Li*[4, 5]

[1]School of Computer Science and Engineering, [2]Sino-French Engineer School, Beihang University, Beijing 100191, China, [3]Department of Physics, Bar Ilan University, Ramat Gan 52900, Israel, [4]School of Reliability and Systems Engineering, Beihang University, Beijing 100191, China, [5]Science and Technology on Reliability and Environmental Engineering Laboratory, Beijing 100191, China.
*daqingl@buaa.edu.cn

01 June 2016

## Abstract

Recently, with the availability of various traffic datasets, human mobility has been studied in different contexts. Researchers attempt to understand the collective behaviors of human movement with respect to the spatio-temporal distribution in traffic dynamics, from which a gravitational scaling law characterizing the relation between the traffic flow, population and distance has been found. However, most studies focus on the integrated properties of gravitational scaling, neglecting its dynamical evolution during different hours of a day. Investigating the hourly traffic flow data of Beijing subway network, based on the hop-count distance of passengers, we find that the scaling exponent of the gravitational law is smaller in Beijing subway system compared to that reported in Seoul subway system. This means that traffic demand in Beijing is much stronger and less sensitive to the travel distance. Furthermore, we analyzed the temporal evolution of the scaling exponents in weekdays and weekends. Our findings may help to understand and improve the traffic congestion control in different subway systems.

# Introduction

In recent years, human mobility features have attracted much research attention [1]. Given the accessibility into different kinds of datasets related to human movement, such as bank notes data[2], locations of mobile phone users[3][4][5][6], check-ins[7][8] and locations of taxis[9][10], human mobility in different aspects have been studied. Human mobility based on transportation networks is also an active field of investigation, including public rail transit network[11] and city road network [12]. According to the spatial and temporal trajectories of human movement, these studies focus on collective mobility distribution in different scales. Among these studies, some attempts have been performed to describe human mobility by means of different scaling laws including gravitational scaling [13].

The gravity model[14], which is analogous to the famous Newton's law of gravity[15] in physics, describes object interactions based on their masses and distance, is widely observed in human mobility studies. For example, Goh et al. [16] analyzed the Seoul subway system and revealed that traffic in subway lines follows gravitational scaling, suggesting that traffic flow divided by the product of masses (total number of passengers) between stations has an inverse power-law dependence with distance. Later, Goh et al.[17] found power law relation of passenger flow strength correlation function as a function of time distance in Seoul bus system, which is probed by the scaling and renormalization analysis of the modified gravity model. In a transportation system, gravitational scaling has been found not only for inner city traffic, but also for intercity flows. Jung et al. [18] observed that intercity traffic flow in Korean highway system could be described with a gravitational scaling law, where the mass of the city is represented by its population. Kwona et al. [19] found also gravitational scaling in the intercity express bus system in Korea, and found that the total bus flow of a city is mainly related to the city's population, with an almost linear dependence. Balcan et al.[20] found gravitational scaling for the interaction in short range commuting flows, where distance scale is up to 300 kilometers and identified different contributions of large and short scale flows to the epidemic spreading. Kaluza et al.[21] classifies the ships traffic into three categories according to ship types and movement patterns in the global cargo network, and investigated how the gravitational scaling law can predict ship movements. Simini et al.[22] introduced a radiation model instead of gravity model, which requires only the information on the population distribution. This model can be applied in areas without detailed mobility information. Except from transportation systems, gravitational scaling is also observed in communication systems. Krings et al.[23] found gravitational scaling for intercity communication intensity between 571 cities in Belgium. Lambiotte et al.[24] found a power-law degree distribution for a communication network constructed by mobile phone customers and used the gravity model to predict the probability that two customers at certain distance have connections. Similar with this phenomena, Goldenberg et al.[25] studied links of users on the Facebook and email communications, and found Zipf's law between probability density of social communications and distance. Kang et al.[26] investigated intercity mobile communication network in China, and found that the communication intensity based on mobile call number or mobile call duration could be described by a gravity model.

Actually, the above studies mostly focus on the static properties of human mobility, while the dynamical evolution of mobility is rarely analyzed. In Goh's study [16] on Seoul subway system, a day was divided into three time zones, namely, morning, afternoon and evening, each time zone lasts 5 to 8 hours. They found 3 similar gravitational scalings for different time zone. Indeed, in a transportation system, the transition between free flow and congestion can be observed in relatively short time scales. Therefore, we propose here to explore and compare the temporal properties of gravitational scaling in peak hours and off-peak hours within shorter time scale windows. In this paper, we study the evolution of traffic flows in Beijing subway network, as shown in FIG. 1 and compare it with that of Seoul. With the data capturing the number of trips from an origin station to a destination station, between all pairs of stations, we study the temporal evolution of traffic scaling in Beijing subway network. Assuming the passengers' sensitivity to number of hops rather than actual distances, we also analyzed the gravitational scaling with respect to the hop-count distance. We find that the daily gravitational scaling for hop-count distance has a scaling exponents approximately 0.4 (for short trips <10 hops). Our results also suggest that passengers are more sensitive to hop-count distance rather than physical distance. These scaling exponents are smaller than those of Seoul, suggesting the Seoul passengers are more sensitive to distances, i.e., try more avoiding large distances compared to Beijing passengers. Furthermore, the temporal evolution of scaling exponents during the day is analyzed. Our analysis also suggests different scaling exponents between weekdays and weekends, which can be associated with passengers' different travel tendencies.

# Results

The gravity model can be used for studying and understanding humans' mobility patterns, which is widely reflected in transportation systems. Traffic gravity model describes the mobility interaction between locations, where two locations' attraction is assumed to scale with the product of their masses (populations) and inversely scale with their distances. For a traffic network composed of $N$ nodes (stations), the gravitational interaction between two nodes (stations) $i$ and $j$ is usually defined as

$$F_{ij} \sim \frac{(M_i M_j)^\beta}{r_{ij}^\alpha}, \qquad (1)$$

where $F_{ij}$ is the non-directed traffic flow between stations $i$ and $j$, which is the sum of flows from $i$ to $j$ and from $j$ to $i$. $M_i$ is the mass of node $i$, which can be represented by the population using this station, i.e., the number of people entering this station and exiting this station during a typical day. $r_{ij}$ is the distance between nodes $i$ and $j$, which can be either physical distance or hop-count distance in the subway network. $\alpha$ and $\beta$ are the scaling exponents of distance and mass respectively.

Here we focus on the subway mobility based on the dataset of the traffic flow in Beijing subway network in 2012. The data captures the number of trips from origin station to destination station (OD flow), which spans three months from January 1st 2012 to May 30th 2012. We mainly calculate the fraction of traffic flow (with respect to total flow in the system) instead of real traffic flow. Due to the schedule of Beijing Subway, we explore the subway flow in the working hours from 5am to midnight. In Beijing subway system, Airport Express is an independent subsystem with payment system of its own, where passengers need to take the Airport Express after getting out from the subway system. Besides, Line FANGSHAN and Line 9 are also not fully connected to the whole network. Therefore, we ignore terminal 2 and terminal 3 of Airport Express, as well as stations in the Line FANGSHAN and Line 9. Excluding these specific stations, there are totally 169 stations in the subway network.

As a major transportation network, the subway system reflects characteristics of intra-urban mobility, especially collective movements. The subway in Beijing has become one of the main infrastructures to support the traffic demand of urban residence. In 2013, the transit rail traffic covered nearly 40% of total public traffic in Beijing. According to the records of daily trips, we show in FIG. 1A sampling of OD flows in a typical day in the year 2012, where only lines above 4000 passengers per day are plotted. For the subway traffic, it is shown in FIG.1 that the flow distribution is not uniform. Significant flow is concentrated in the center of Beijing, and a few large flow trips are distributed mainly among lines 1, 2, 5, 10 and 13.

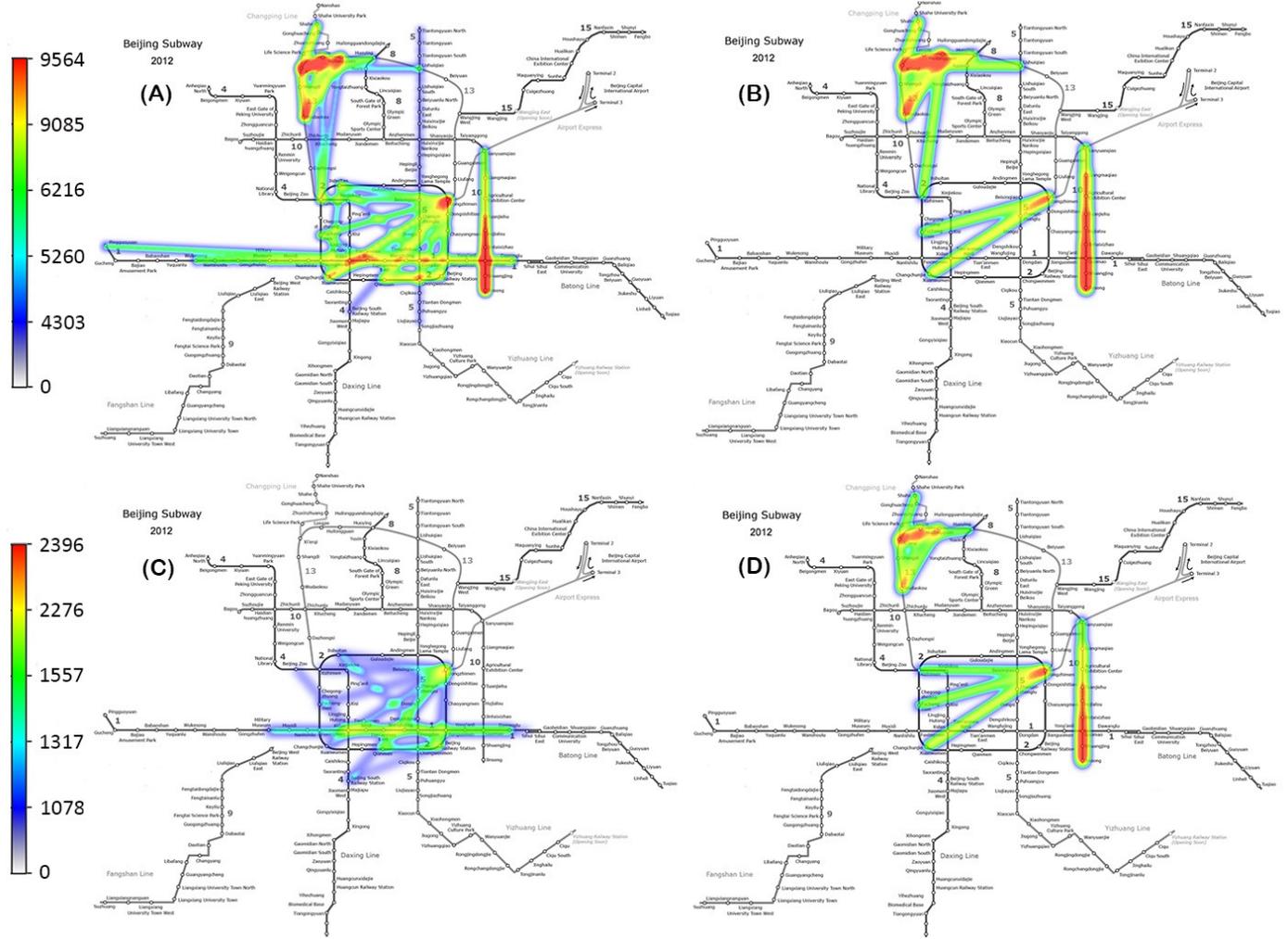

FIG. 1 The heat map for traffic flow in Beijing in May 14$^{th}$ 2012. (A) The daily non-directed flow between two stations is plotted, the colors represent the total number of travelers in one typical day. In the figure, only flows between two stations over 4000 passengers are plotted including 39 pairs of stations. For example, the total trips between Station TIANTONGYUAN North and Station LIUJIAYAO during this day is 4001 (blue). The maximal flow is 9564 trips, which is between CHANGCHUNJIE Station and DONGZHIMEN Station (red). The traffic flow every two hours is plotted in (B) from 7:00 to 9:00, (C) from 11:00 to 13:00 and (D) from 17:00 to 19:00. For these three time ranges, the top 15 flows between pairs of stations are shown in each figure using the same color bar. The maximal flow from 7:00 to 9:00 is 2396 trips, which is between XIERQI Station and HUILONGGUAN Station (red).

To identify the flow distribution of trips with respect to distance and to capture a general insight on the collective mobility, we analyze the relation between the relative daily passenger flow (between pairs of stations) and distance. We can see in FIG. 2(A), for distances below 10 kilometers, the average daily flow is almost not sensitive to the physical distance. In this paper, physical distance measures the geographic distance between two stations using their geographic coordinates of latitude and longitude. For pairs of stations having distances above 10 kilometers, the daily flow begins to decrease with increasing distance. This insensitivity (below 10 kilometers) may come from the combined effect of high commuting demand and low ticket price (2 Yuan per trip≈0.3 dollar).

The exploration of relation between passenger flow and distance cannot reflect fully the scaling law in gravity model. Actually, in the transportation system, the passenger flow is determined not only by the distance between two stations, but also by the mass (population) of each station. In a subway system, the mass of a station could be represented by the sum of daily inflow and outflow through this station, during the day. Here we calculate the mass of station by taking the average daily inflow and outflow during the five months in 2012, and use the fraction of mass in each station, with respect to total mass in the system. To study the mass exponent, $\beta$, in the gravity model, we explore the relation between daily flow and mass. FIG. 2(B) shows the positive relation between normalized daily flow and product of two stations' masses in a double logarithm diagram. Fitting the data with least square method, the exponent is found close to 1, which is consistent with the assumption of entropy maximization[27].

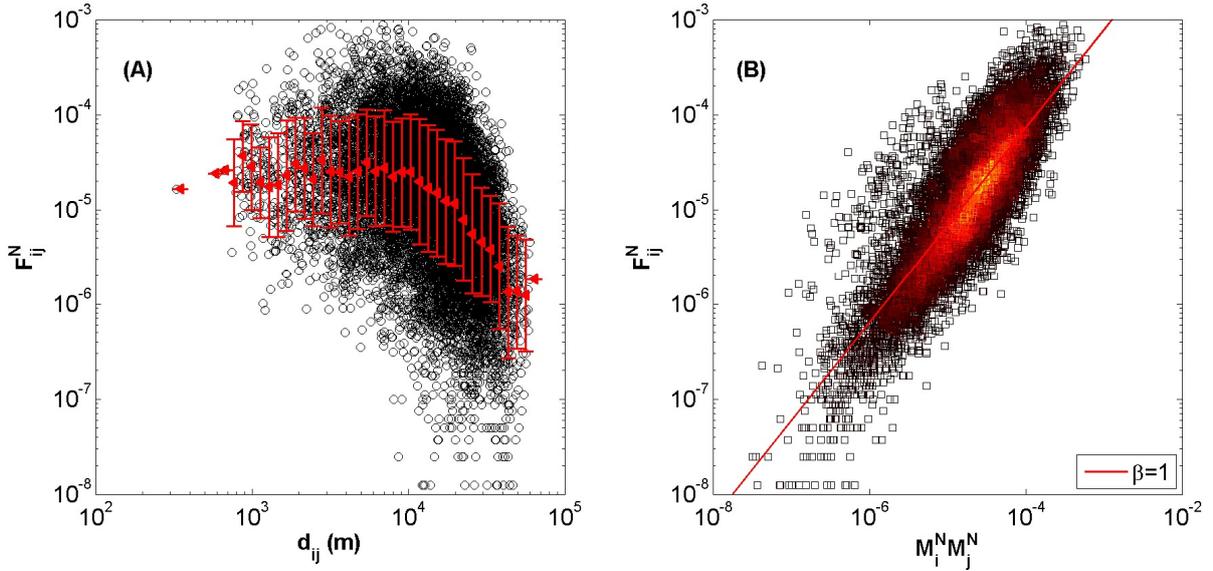

FIG. 2  (A) Relation between daily flow and their physical distance in Beijing subway during 2012. The daily flow between two stations is the average daily flow of ten weekdays (from May 7th to May 12th and from May 14th to May 18th). The normalized daily flow counts the relative flow (with respect to total) between all pairs of stations in the network (See Eq. (1) and Eq. (5) in SI). The red triangles are the average relative flow in a logarithmic binned distance window with standard deviation calculated over the 10 working days. (B) Relation between the relative daily flow and the product of two stations' relative masses. The mass is the total daily inflow and outflow for a given station averaged over five months from January to May (See Eq. (6) in SI). The relative mass is the fraction of mass compared to the total mass between all the stations. The slope $\beta = 1$ supports the linear relation of the flow with the product of masses in Eq. (1).

To better determine the scaling law between flow and distance, we use the definition of reduced flow [16] between nodes $i$ and $j$, defined as,

$$f_{ij} = \frac{F_{ij}^N}{(M_i^N M_j^N)^\beta} \,. \tag{2}$$

According to FIG.2 (B), $\beta$ is equal to 1. To investigate the scaling exponent $\alpha$, we explore the relation between reduced flow, $f_{ij}$, and distance between stations. Using Eqs. (1) and (2) we expect that in some range of scales $f$ can be approximated by

$$f \sim r^{-\alpha} \,, \tag{3}$$

where $r$ is physical distance $d$ or chemical distance $l$ (number of hops). As shown in FIG.3, similar to $F_{ij}^N$, $f$ is again not sensitive to the physical distance $d$ for $d$ smaller than 10 kilometers. Note that this result is different from the scaling behavior found in Seoul subway system [16], where it is found that the reduced flow increases slightly with distance in a short time (distance) range, from zero to about 11 minutes, then the reduced flow decreases with distance. Note that the distance in Seoul subway is measured in time units. In contrast, for Beijing subway network, the reduced flow remains almost constant for physical distance below 10 kilometers, where 10 kilometers correspond to travel time of 30 minutes on average.

Considering the particular case of the rail transit system, the movement trajectory is under the restriction of the railway pathway. Thus, as discussed above, in addition to the physical distance, we also consider the chemical distance metric, i.e., the hop-count distance (number of stops between stations) in the subway network as a distance measure in the relation between traffic flow and distance. Given the short travel time between stations, the hop-count distance can represent the travel time of each trip. Indeed, the scaling exponent based on hop-count distance is found larger than that of physical distance, as shown in FIG.3 (B), indicating that the passengers are more sensitive to length of trips based on number of hops rather than that based on physical distance between two stations. Here the hop-count distance is assumed proportional to the travel time, considering similar distance between subway stations. Comparing the scaling exponent 0.4 found in Beijing and 1.6 found in Seoul, it is shown that passengers in Beijing subway system have much stronger demand for subway traffic.

The difference of scaling exponent between hop-count distance and physical distance is comprehensible. When one is choosing the traveling route in subway system, hop-count distance is probably put on more weight than physical distance, given the high speed of trains in the subway. Note that the two measures of distance have a heterogeneous corresponding relation, which may be the result of different scaling exponent. For some pairs of stations having large hop-count distance, the distribution of corresponding physical distances can be broad (see SI for more details).

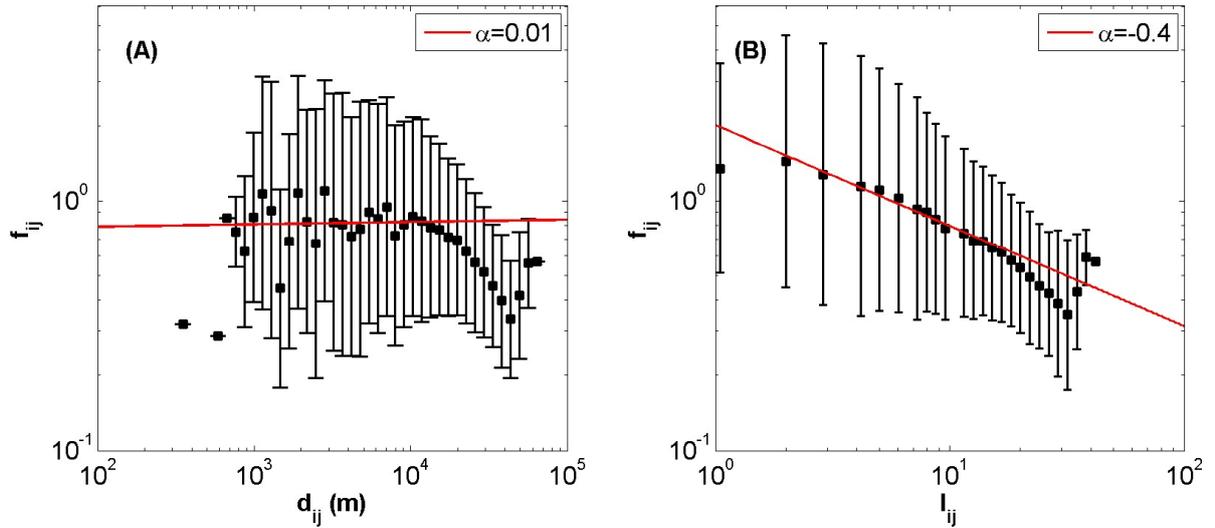

FIG. 3 Relations between daily reduced flow and (A) physical distance (B) hop-count distance. The reduced relative flow between two stations is the flow divided by the product of their masses (see Eq. (2)). The black squares are the average reduced flow of ten weekdays from May 7$^{th}$ to 18$^{th}$ in 2012 in a log binned distance window. For physical distance, the minimal distance of two stations is 0.328 kilometers between NANLISHILU Station and FUXINGMEN Station, the maximal is 60.113 kilometers between TIANGONGYUAN Station and FENGBO Station. There are only few pairs of stations, whose distance is smaller than 1 kilometer or larger than 50 kilometers. For Beijing subway network in 2012, the maximal hop-count distance is 40, between TIANGONGYUAN Station and FENGBO Station. The minimal hop-count distance is 1, which is between adjacent stations.

While many researchers investigated the integrated properties for gravitational mobility scaling, little is known on temporal evolution of gravitational scaling in short time scales. To study the evolution of gravitational scaling, we analyze the scaling relation between flows and hop-count distance every two hours in three months in 2012. We define reduced flow as the reduced flow in two hours between two stations. The reduced flow is calculated by the fraction of traffic flow in both directions in these two hours using Eq. (2). The traffic flow in two hours counts the number of trips according to time entering and exiting the station. The relative hourly mass of each station is defined as the sum of relative flow between this station and all other stations in two hours. We divide a day from 5:00 to 23:00 into nine two-hour sections, neglecting the flow from 23:00 to 24:00 due to its low values. Note that there are less than 1% of pairs that take more than two hours to reach from one station to the other. Therefore, two hours period can help us to explore the traffic flows' temporal evolution.

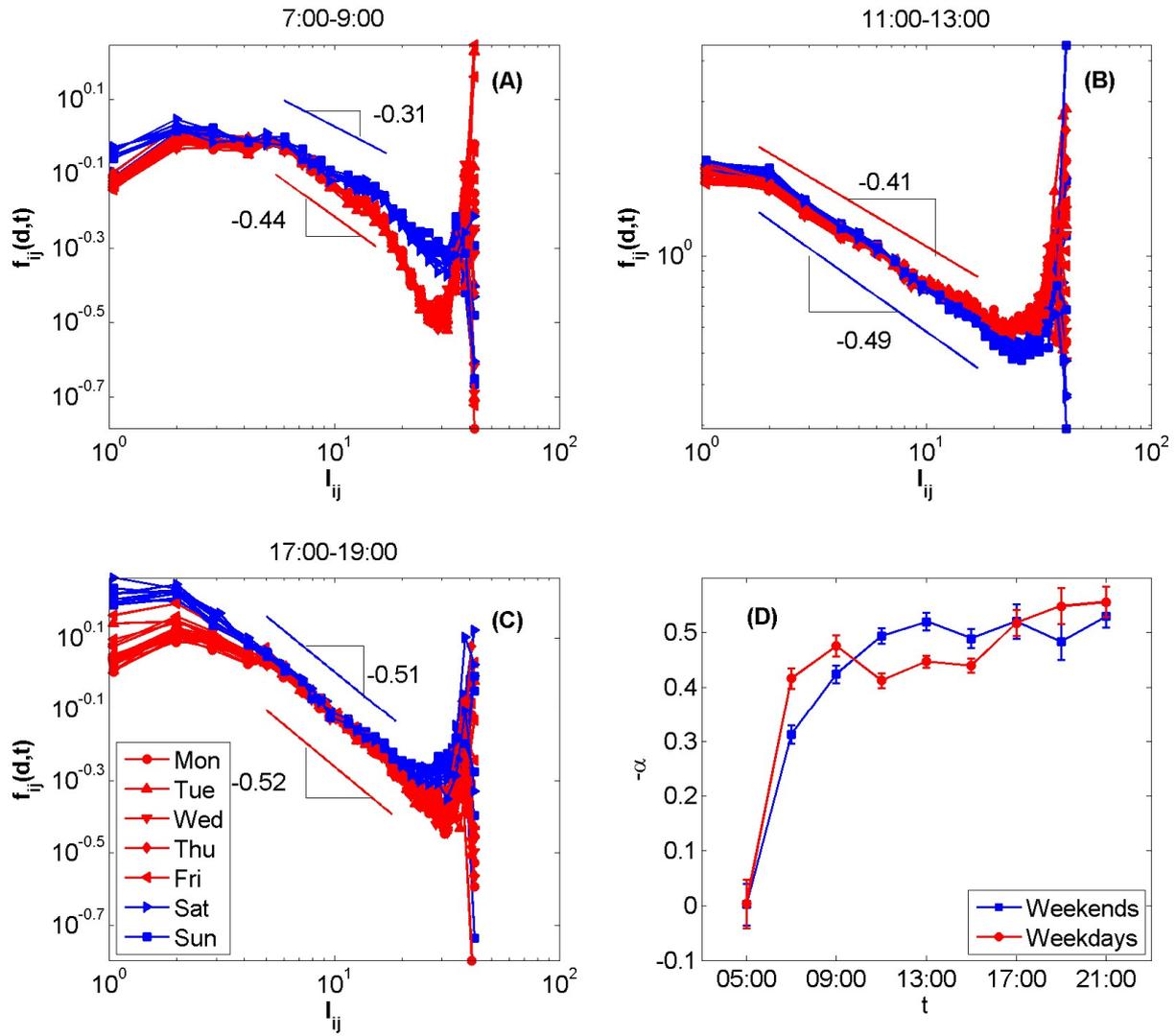

FIG. 4 Relations between hourly reduced flow and hop-count distance in weekdays and weekends from March 19th to May 21st 2012 during (A) 7:00-9:00 (B) 11:00-13:00 (C) 17:00-19:00. The symbols are the average hourly reduced flow in a log binned distance window. The data includes 5 weeks (35 days) in these three months except holidays from March 19th to May 21st. The red symbols represent weekdays and the blue ones represent weekends. (D) The averaged scaling exponents of reduced flow with error bars as a function of time in weekends and weekdays.

There are mainly four categories of traffic in these nine sections of two-hours. For the early morning (from 5:00 to 7:00), most passengers take subway for work in a long journey (see SI for more details). The second category is the peak hours for commuting, including morning peak (from 7:00 to 9:00) and evening peak (from 17:00 to 19:00), as shown in FIG. 4. The third category is noon time period (from 11:00 to 13:00). The last category is the periods after 19:00, during which many passengers take journey for home (see SI for more details). Our analysis suggests that gravitational scaling laws can be different in these different time windows, and also different in the weekdays and weekends as seen in FIG.4.

For weekdays, the time interval from 7:00 to 9:00 is defined as morning peak period. As shown in FIG. 4(A), the hourly flow does not follow well a gravitational scaling. Only trips with medium hop-count distance seem to follow a scaling with exponent close to 0.44. For trips less than 5 hops, hourly flow is almost independent on hop-count distance. This phenomenon is probably due to the commuting effect in weekdays, which results in a stable (independent of distance) demand for subway traffic for short distances. In the mid-day period in FIG.4 (B), hourly flow in weekdays shows well a gravitational scaling behavior with an exponent 0.41, which is slightly lower compared to those in morning and evening peaks. Different from morning peak hours, 17:00 to 19:00 are evening peak hours, with various traffic demand ranging from social activities to entertainment. The exponent is higher than that of morning peak and mid-day period, representing the higher sensitivity of passengers to hop-count distance. Therefore, a larger scaling exponent (0.52) is found during this interval. In weekends, in the morning peak, the scaling exponent is about 0.31. This low value exponent indicates passengers' insensitivity to distance, probably due to vacation trips. In the mid-day period, the gravitational scaling exponent is about 0.49. In the evening peak, the scaling exponent is 0.51. In the afternoon and evening in weekends, the scaling exponents are similar. In these periods, people probably tend to make shorter trips considering the significant traffic congestion and relevant trip costs.

FIG.4 (D) compares the evolution of gravitational scaling exponent during a day of weekdays and weekends. Three different stages can be found: at 5:00-7:00, the traffic demand is rigid and the exponent is almost zero; for the morning and noon, the exponent is around 0.4, while the exponent increases in the evening period, suggesting that people tend to take shorter trips during this time window. From the early morning to the late evening, a tendency can be clearly observed that the traffic demand in subway is relaxing as a result of loosen requirement, when passengers have more options to choose in the evening.

It is interesting to note that the reduced normalized flow has usually a local maximum for large hop-count distance in different periods of a day. These trips are mostly the result of the origin being at the border of the subway network, and the end being in another border station (further details can be found in SI). As the only branch into the rural neighborhood, these border stations can have larger traffic demand due to serving bigger areas, compared to other stations.

# Conclusion

We investigated the gravitational scaling of Beijing subway system based on the traffic flow data in 2012. Different scaling performances based on physical distance and hop-count distance are observed. The smaller scaling exponent in Beijing subway network is revealed compared to Seoul subway system, which indicates the weak distance sensibility of passengers in Beijing. The difference of scaling exponent between Beijing subway system and Seoul subway system may also be caused by the spatial distribution of functional location and pricing policy [28], in a long term viewpoint. Besides, we analyzed dynamical performance of scaling exponent in hourly time scale and found different scaling exponents in weekdays and weekends. These differences may come from dynamical states of the whole Beijing traffic, where the lower scaling exponent can be observed due to the presence of severe traffic congestion and high demand for commuting. The gravitational scaling can help to understand the macroscopic traffic properties in the subway systems, which is essential for the design and optimization of traffic efficiency.